\documentstyle[twocolumn,prc,aps,epsfig]{revtex}
\input epsf
\newcommand{\be}{\begin{eqnarray}}
\newcommand{\ee}{\end{eqnarray}}

 \newcommand{\gsim}{\mathrel{\hbox{\rlap{\lower.55ex \hbox {$\sim$}}
                   \kern-.3em \raise.4ex \hbox{$>$}}}}
\newcommand{\lsim}{\mathrel{\hbox{\rlap{\lower.55ex \hbox {$\sim$}}
                   \kern-.3em \raise.4ex \hbox{$<$}}}}

\def\roughly#1{\mathrel{\raise.3ex\hbox{$#1$\kern-.75em%
\lower1ex\hbox{$\sim$}}}}
\def\lsim{\roughly<}
\def\gsim{\roughly>}

\def\la{{\Big<}}
\def\ra{{\Big>}}

\begin{document}

\twocolumn[\hsize\textwidth\columnwidth\hsize\csname @twocolumnfalse\endcsname

\title {
{Instanton-induced Inelastic Collisions in QCD}}
\author { Maciek A. Nowak$^{a,b}$, Edward V. Shuryak$^b$ and Ismail Zahed$^b$}
\address {$^a$ M. Smoluchowski Institute of Physics, Jagellonian
University, Cracow, Poland\\
$^b$ Department of Physics and Astronomy, State University of New York,
     Stony Brook, NY 11794
}
\date{\today}
\maketitle
\begin{abstract}
We show that the instanton-induced inelastic processes, leading to
multi-gluon production in 
high-energy parton-parton scattering, are considerably enhanced 
 over the quasi-elastic ones, by a factor of $100$.
 The basic instanton-induced 
inelastic contribution cause the parton-parton cross section
to increase as  ${\rm ln}\,{s}$, and their  Poisson resummation 
 in hadron-hadron scattering yield 
a regge-type cross section. The pomeron slope and intercept
due to instanton-induced contributions are evaluated. We show 
that the small intercept is due to the diluteness of the instantons
in the QCD vacuum, while the small slope is related to the smallness of 
the instanton sizes.

\end{abstract}
\vspace{0.1in}
]
\newpage

\section{Introduction}\label{intro}

QCD instantons~\cite{BPST,CDG} play an important role 
in the composition of the vacuum and its hadronic
excitations~\cite{SS_98}. This viewpoint is strongly
supported by detailed lattice simulations~\cite{SS_98}.
Naturally most hadronic substructure, whether in the form
of constituent quarks or gluons, should also be important 
for hadronic reactions at high energies. The problem 
in translating vacuum physics to high energy 
scattering has been strongly limited by technical issues, 
an important one being the Euclidean nature of instanton physics 
and the inherent 
light-cone character of high-energy kinematics. As a result,
the theory of high energy processes remains mostly
perturbative, as best illustrated by the BFKL ladder 
re-summation~\cite{BFKL} in the hard regime,
with exchange momenta much larger than 1 GeV. 

This not withstanding, a rich  pomeron phenomenology has been developed 
prior and through  QCD. We will not be able to render justice here to
all relevant papers. Particularly important for us
is the formulation based on the eikonal expansion for the high
energy parton-parton cross section,  originally suggested by 
Nachtmann~\cite{NACHTMANN}.  Similar expressions for 
structure functions were also suggested by Muller \cite{Muller}.
Ideas using instanton effects for  high energy QCD processes were
also recently discussed, for dipole cross sections in~\cite{ES00},
and the soft pomeron problem in~\cite{KL,KKL} (KKL).

Recently, we have suggested a non-perturbative approach to high energy 
scattering using instantons~\cite{sz00}.
The eikonalized near-forward parton-parton scattering 
amplitude was reduced to a pertinent correlation function of two
(or more) straight
Wilson lines, which were analyzed in Euclidean space using instantons.
The lines at an arbitrary
angle $\theta$,  which is then analytically continued to Minkowski space by 
the trick $y=-i\theta$ where $y$ is the Minkowski rapidity. A similar
construction was applied recently to nonperturbative parton-parton scattering in
supersymmetric theories using the AdS/CFT correspondence~\cite{SUPER},
where on the boundary large $N$ instantons are expected to saturate
exactly the diffractive cross section. (Incidentally, most of the arguments
to follow can be checked exactly in these theories using  instanton
calculus).  In the instanton
field the parton Wilson lines involve multi-gluon exchange as depicted
in Fig.~\ref{fig_diag}, with no need for initial and final state
multi-gluon resummation. Our analysis has shown that the
cross section for ``{quasi-elastic}'' (color-transfer)
parton-parton and dipole-dipole scattering is constant at large
$\sqrt{s}$.

In this paper we extend our original analysis to ``truly inelastic''
parton-parton scattering amplitudes,  with prompt parton production.  
Such partonic processes have new particle lines crossing the unitarity 
cut as shown in the lower-left  part of  Fig.~\ref{fig_diag}.
Similar processes have been studied in the early 90's, in the context
of baryon-number violation in the electroweak
theory\cite{weakinst}. In the latter it was shown that multiple gluon production
 can also be calculated  semi-classically, through
interacting instanton-antiinstanton configurations or
stream-lines\cite{streamline} which interpolate between
a well separated instanton-antiinstanton and the vacuum.
By combining the semi-classical treatment of multiple incoming gluons,
as done in \cite{sz00}, with the  stream-line-based treatment of
multiple outgoing gluons, we  can  completely bypass the
perturbative expansion, as indicated in the lower-right part of
Fig.~{\ref{fig_diag}. Contacts with perturbation theory
follow by expanding the instanton contributions in powers of the field
in the weak-field limit.

In this paper we will not develop a quantitative theory
of hadronic collisions (as that require further modeling)
and consider only the basic process involving inelastic quark-quark 
scattering. For clarity,
it is important that the underlying assumptions 
in our analysis be spelled out from the onset. Throughout,
the scattering 
processes are understood to undergo three sequential stages:\\
\indent{\bf i. initial stage:} Partons are initially described
by some wave-function in a fixed frame (say the CM), depending on
their transverse
momenta and rapidities~\footnote{The parton model is of course frame
and scale specific:
 partons which belong to a wave function of one colliding hadron
 in a given frame, can  belong to another hadron wave function 
 in another frame. The normalization scale is basically given by the
inverse instanton size.}. The {\em through-going partons} 
should be formed outside of the instanton of mean size $\rho_0$
with  $k_\perp^2/\sqrt{s}\leq 1/\rho_0$, while the {\em wee partons} 
(the opposite condition) are not in the wave function but included in the cross section. 
 The formers are assumed to move along the eikonalized straight 
 Wilson lines, while the latters are part of the process.\\
\indent{\bf ii. prompt stage:} In it the incoming partons pass each other.
Color of through-going partons could be changed 
(quasi-elastic), or new partons/hadrons could appear (inelastic).
In analogy with the perturbative treatment, confinement is ignored 
at this stage, since the passage time
is short (of order $1/\sqrt{s}$). All partons interact with instantons.\\
\indent{\bf iii. final stage:} In it all produced partons fly away, 
some dragging longitudinal color strings of matching color. String
breaking happens with probability one, thus cross sections are not affected.
These eventually  produce the physical final states
with multiple hadrons.\\
\noindent All partonic amplitudes to be assessed, will
take place in the prompt regime. We assume that by duality, the total 
partonic cross sections match the hadronic cross sections.

In section 2, we give an overview of the salient
physical points of this work. In section 3, we give a brief
account of the instanton-induced quasi-elastic (color-transfer),
parton-parton scattering amplitude as reported in~\cite{sz00}. 
We also report on novel issues regarding the character of the
weak-field limit in light of perturbation theory, as well as the 
absence of odderons in instanton-induced processes.
In section 4  we argue that the full inelastic
contribution to the parton-parton scattering amplitude follows
semi-classically from the QCD streamline configuration. The result is
a remarkable enhancement in the inelastic scattering amplitude,
limited only by the unitarity bound. In section 5
we show that
a statistical resummation of the enhanced pair cross sections yield
a reggeized hadron-hadron scattering cross section. The pomeron
intercept and slope are sensitive to the QCD vacuum parameters,
and as it turns out, even to the instanton shapes. 
In section 6 we draw a parallel between the Weizsacker-Williams 
approximation to inelastic scattering and the weak field
limit of the semi-classical analysis, and discuss which
instanton-induced form-factors are 
the pertinent ones. In section 7 we discuss additional contributions
to the diffractive process not retained in our analysis.
Our conclusions and outlook are summarized in section 8.

\section{Physical Highlights}

The ubiquitous character of the instantons in the nonperturbative
QCD vacuum, as now established both theoretically and numerically, 
leads naturally to their importance in partonic scattering during 
the collision time  (prompt stage). In particular, instantons prove
essential for discriminating perturbative from nonperturbative effects.
Indeed, whenever instanton effects are included (even in lowest order)
one can often locate the non-perturbative boundary in perturbation
theory, and even makes meaningful predictions a little beyond
it~\cite{SS_98}. Although in QCD instanton-induced effects have a small
appearance probability (density) $n_0\approx e^{-2\pi/\alpha_s}$, they
correspond to strong (classical) fields $A\approx 1/g$. Hence, any interaction with
a parton of charge $g$ is $gA\approx 1$ which is independent of the 
charge. In contrast to perturbation theory, there is no additional
penalty  for adding partons to the amplitude. Therefore,
the instanton-induced amplitudes overcome the perturbative amplitudes
at high enough order. Indeed, recently we have suggested~\cite{sz00}
that instantons would dominate collisions with multiple 
color exchanges between partons, leading to the higher observed hadronic 
multiplicities.
\begin{figure}[t]
\vskip 0.2in
\includegraphics[width=2.3in]{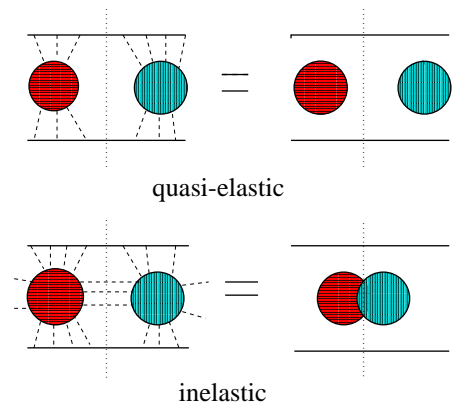}
\vskip 0.2in
\caption[]{
 \label{fig_diag}
Schematic representation of  the amplitude squared, with (without) gluon
lines  are  shown in the left (right) side of the figure. The dotted
vertical line is the unitarity cut.
The upper panel illustrates the  quasi-elastic (at the parton level)
amplitudes where only color is exchanged as detailed
in~\cite{sz00}. The lower panel depicts inelastic processes in which
some gluons cross the unitarity cut, and some gluons are absorbed in
the initial stage.
}
\end{figure}

The important question regarding the transition from the perturbative
regime to the instanton dominated (semiclassical) regime in
parton-parton collisions, depends critically on the numerical parameters
characterizing the QCD vacuum. There are essentially two key 
parameters~\cite{Shu_82} (see also \cite{sz00} for some details). The
instanton (plus anti-instanton) density
$n_0\approx 1 \,/ {\rm fm}^{4}$ and the mean instanton
size $\rho_0\approx 1/3 \, {\rm fm}$ yield
the dimensionless diluteness $\kappa_0=n_0\rho_0^4\approx 0.01$ of the 
instanton vacuum~\footnote{For comparison 
we note that in the electroweak theory the diluteness 
factor is $10^{-84}$ despite the fact that the coupling is only three
times smaller than in QCD.}. The mean instanton action is $S_0=2\pi/\alpha_s\approx 10 - 15$.
Each time an instanton is inserted it costs a small factor $\kappa_0$.
However, there are no coupling constants, and so each time we compare
the results with their perturbative counterparts we get
powers of the large action $S_0$, with a net gain per gluon involved. 
Numerically the
instanton-induced effects should dominate the perturbative effects
from third order and on ($\kappa_0\,S_0^2 \approx 1$), while
they comparable to second order. Of course this argument is too naive,
there maybe other factors and so on, but we believe the argument
captures the main reason why instanton-induced processes  
dominate the inelastic parton-parton cross sections.

The quasi-elastic (QE) parton-parton scattering cross 
section \cite{sz00}, produces a cross section of order
\be 
\sigma_{QE}\approx \pi \rho_0^2\,\,\kappa_0^2 \,\,,
\ee
which is small in comparison to the one-gluon exchange
(OGE) result at the same scale\footnote{The infrared 
sensitivity in OGE is cutoff by $ 1/\sqrt{-t}\approx \rho_*$.} 
\be \label{eq_OGE}
\sigma_{OGE} \approx \pi \rho_*^2\,\,({\alpha_s /\pi})^2\,\,. 
\ee
Below, we show that the instanton-induced parton-parton
inelastic cross section is significantly enhanced
\be 
\sigma \approx \pi \rho_0^2\,\,\kappa_0\,\,,
\ee
by one power of the diluteness factor $\kappa_0$, even though
it produces about $S_0\approx 10-15$ gluons. The perturbative 
contributions to the inelastic cross section are suppressed.
Indeed, a {\it one-gluon} production yields
\footnote{Both the instanton-induced and perturbative amplitudes 
yield ${\rm ln}\,s$ enhancements that are summed.}
\be 
\sigma_{ gg \rightarrow ggg } \approx \pi\, \rho_*^2 \, ({\alpha_s /
\pi})^3 \,\,.
\ee
As a result, the instanton-induced effects 
dominate the perturbative contributions in the growing part of the
inelastic cross section.

The  distribution over the invariant mass $Q$ of the produced
gluons, deposited by the wee
incoming partons onto the instanton, 
 will be calculated in a specific unitarization model. In
the weak field limit the growth in the inelastic cross section
$gg\rightarrow {\rm any}$ is captured by the ``holy-grail function''
\be
\sigma_{gg\rightarrow {\rm any}} (Q)\approx  e^{F(Q/Q_s)}\,\,,
\label{holy} 
\ee
 studied in the baryon number violating processes in the
standard model~\cite{weakinst}. 
It peaks around the so called  sphaleron
energy, which in QCD is given
by the mean instanton size $\rho_0$
\be
E_s=Q_s/\rho_0 \approx 3\pi/(4\rho_0\alpha_s)\approx 2 \, {\rm GeV}\,\,,
\label{ESTIMATE}
\ee
for $\rho_0\approx 1/3 \,{\rm fm}$ and $\alpha_s\approx 1/2$.
Below this sphaleron energy the cross section
is small but
 rapidly growing. As  shown by Khoze and Ringwald\cite{weakinst},
this growth can be technically attributed to a moving saddle point, 
reducing the relative distance between an instanton and its conjugate
antiinstanton, thereby decreasing their initial action of $2S_0$.
Reliable calculation can be carried out in this region. The increase
in the inelastic amplitude is stopped by unitarity constraints, as
suggested by Zakharov~\cite{ZAKHAROV}, precisely when the total
instanton
 action
is reduced from 2$S_0$  to $S_0$.
The specifics of the unitarization process to
be discussed below will follow on the qualitative arguments suggested
by Maggiore and Shifman\cite{MS}.
Aside from  technicalities,
the physical meaning of their arguments is simple:  if there is enough
energy for the system  to reach the top of the barrier (the sphaleron),
its consecutive decay follows with unit probability. All what is needed is
that the ``wee partons'' (the field of the through-moving hard partons)
can encounter an instanton and deposit about 2 GeV of invariant mass.
The tunneling probability at low energy  on the other hand
can be considered as a product of the amplitude to get {\it on} and 
{\it off} the barrier: hence the instanton amplitude appears twice.
In QCD this enhancement of the inelastic processes
with multi-gluon production
 amounts to the gain $\kappa_0^2\rightarrow \kappa_0$
which is an enhancement by a factor of the order of a 100 in the
inelastic cross section relative to the quasi-elastic one
\footnote{The analogous electroweak instanton-induced
 cross section is increased by
about 85 orders 
of magnitude due to the same enhancement:  but   it still remains
 far below any observable rate because one power of small diluteness is
 there.}.

Although we do not use the concept  of t-channel gluon exchanges
(as they are summed into the eikonalized phases), some
of its features remarkably survive. Already
in the quasi-elastic process only the octet color exchange survives
the high-energy limit ~\cite{sz00}. The same feature carries over to
the inelastic processes we now consider  where only the octet
color is transmitted from each parton line. Of course they should be
in the same SU(2) subgroup to interact with a given instanton.
In general, these restrictions on the possible color representations
disagree with the exponentiation of multi-gluon production, leading to
the semiclassical theory we use. In a way, this resembles the trade
between using canonical as opposed to  micro-canonical (more accurate)
ensemble in statistical mechanics. Presumably our assessment of the
total cross section is still good, since the number of produced gluons
is large, $N_g\approx S_0\approx 10-15$.

The questions regarding the ``sphaleron decay modes''
(their decomposition into  various channels) will be discussed elsewhere.
We note that in recent phenomenological studies~\cite{KL,ES00} of
high-energy scattering only colorless channels (the rangs of ladders)  were
considered,  e.g. $0^{++}$ (scalar glueball) and $\pi\pi$ (scalar).
Although these states may contribute to the ``sphaleron decay modes'',
especially in light of the closeness of the $0^{++}$ mass to the sphaleron 
energy (\ref{ESTIMATE}), we expect on general grounds additional
contributions involving colored states as well.

Finally, we will show how the logarithmic growth in the instanton-induced
parton-parton scattering amplitude takes place. Empirically, the growth 
is fitted at about $s\approx 100$ GeV$^{2}$. In this regime, the number
of through-going effective partons can still be considered small and
fixed with $qqq$ for the nucleon and $\bar q q$ for mesons and photons
(plus possibly some ``primordial'' glue from the QCD strings). At higher
energies  (although well below the Froissart bound) the power growth
takes place, with possibly several effects contributing to it.  The first 
and the simplest contribution (to be discussed in this work) is that since the 
elementary cross section grows, more parton-parton interaction takes place. 
Since the hadron size is large compared to that of the instanton 
$R_H^2/\rho_0^2\approx 10 \gg 1$, we think that 
double, triple etc parton collisions will take place in a
 statistically independent way (Poisson). 
A straightforward resummation of the compounded probabilities  yields a 
total hadronic cross section that approximately reggeizes
\be
\sigma_H (s,t) \approx \pi\, R_H^2 \, s^{\Delta (t)}\,\,,
\ee
with $\Delta (t)$ of order $\kappa_0$.  In a way, we may think of
$\Delta (t)$ as the square of the instanton-induced  form factor.
The second contribution is a re-interaction of one of the produced gluons,
leading to {\em instanton ladders} as considered in KKL~\cite{KKL}.
The third and final contribution may stem from rescattering through the
``primordial'' partons at small $x$ in the wave function, 
even at the low normalization point under consideration. These effects 
are not included in our ``wee'' partons which are separated in the
transverse plane by a distance larger than $\rho_0$ from all others.

\section{ Quasi-Elastic Scattering}

In this section we will recall the quasi-elastic results derived
in~\cite{sz00}. This will help us streamline the notation 
and facilitate the comparison with the inelastic results to follow. We
also discuss issues related to instantons in the weak-field limit,
and show that instanton-induced processes do not discriminate 
between C-even and C-odd effects in the t-channel, despite their 
multi-gluon structure.

\subsection{The Eikonal Approximation }
Using the eikonal approximation and LSZ reduction,
the scattering amplitude ${\cal T}$ for quark-quark
scattering reads~\cite{NACHTMANN}
\begin{eqnarray}
&&{\cal T}_{AB,CD} (s,t) \approx
-2is \int d^2b\,\, e^{iq_{\perp}\cdot b}\nonumber\\&&\times
\la \left({\bf W}_1 (b) -{\bf 1}\right)_{AC}
\left( {\bf W}_2 (0) -{\bf 1}\right)_{BD}\ra\,\,,
\label{4}
\end{eqnarray}
where
\be
{\bf W}_{1,2}(b)= {\bf P}_c {\rm exp}\left(ig\int_{-\infty}^{+\infty}\,d\tau\,
A(b+v_{1,2}\tau)\cdot v_{1,2}\right)\,\,.
\label{5}
\ee
The 2-dimensional integral in (\ref{4}) is over the impact parameter $b$
with $t=-q_{\perp}^2$, and  the averaging is over the gauge configurations
using the QCD action.  AB and CD are the incoming and outgoing color and spin
of the quarks.

In Euclidean geometry, the kinematics is fixed by noting that
the Lorenz contraction factor translates to
\begin{equation}
{\rm cosh}\,y = \frac 1{\sqrt{1-v^2}} =\frac s{2m^2}-1\rightarrow {\rm cos}\,\theta\,\,.
\label{LO1}
\end{equation}
Scattering at high-energy in Minkowski geometry follows from scattering in
Euclidean geometry by analytically continuing $\theta\rightarrow -iy$
in the regime $y\approx \,{\rm log}\, (s/m^2)\gg 1$.
It is sufficient to analyze the scattering for
$p_1/m=\,(1,0,0_\perp)$, $p_2/m=\,
({\rm cos}\theta\,,-{\rm sin}\theta,\,0_\perp)$, $q=(0,0,q_\perp)$
and $b=(0,0,b_\perp)$.
The Minkowski scattering amplitude at high-energy
can be altogether continued to Euclidean geometry through
\begin{eqnarray}
&&{\cal T}_{AB,CD} (\theta, q) \approx
4m^2\,{\rm sin}\,\theta \int d^2b\,\, e^{iq_{\perp}\cdot b}
\nonumber\\&&\times
\la \left({\bf W}(\theta, b) -{\bf 1}\right)_{AC}
\left( {\bf W}(0,0) -{\bf 1}\right)_{BD}\ra\,\,,
\label{4X}
\ee
where
\be
{\bf W}(b,\theta)= {\bf P}_c {\rm exp}\left(ig\int_\theta\,d\tau\,
A(b+v\tau)\cdot v\right)\,\,,
\label{5X}
\ee
with $v=p/m$. The line integral in (\ref{5X}) is over a straight-line
sloped at an angle $\theta$ away from the vertical. Corrections to
the eikonal approximation will be discussed in the next section.

\subsection{Quasi-Elastic Amplitude}

At large $\sqrt{s}$ the one-instanton
contribution to the color-elastic parton-parton scattering amplitude drops as
$1/\sqrt{s}$~\cite{sz00}. However, the two-instanton contribution to the
color inelastic part survives~\cite{sz00}. To set up the notation, consider
the untraced and tilted Wilson line in the one-instanton background
\begin{equation}
{\bf W} (\theta, b) = {\rm cos}\,\alpha -i\tau\cdot\hat{n}\,{\rm sin}\,\alpha\,\,,
\label{QQI1}
\end{equation}
where
\begin{equation}
n^a ={\bf R}^{ab}\,\eta^b_{\mu\nu}\,\dot{x}_\mu (z-b)_\nu ={\bf R}^{ab}\,{\bf n}^b\,\,,
\label{QQI2}
\end{equation}
and $\alpha=\pi\gamma/\sqrt{\gamma^2+\rho^2}$ with
\begin{eqnarray}
\gamma^2=&&n\cdot n={\bf n}\cdot{\bf n}\nonumber\\=&&
(z_4{\rm sin}\,\theta - z_3 {\rm cos}\,\theta )^2 + (b-z_\perp)^2\,\,.
\label{QQI3}
\end{eqnarray}
The one-instanton contribution to the scattering amplitude (\ref{4X})
reads
\begin{eqnarray}
&&{\cal T}_{AB,CD} (\theta, q) \approx
{\rm sin}\,\theta\, \int d^2b\,\, e^{iq_{\perp}\cdot b}\,\,\int \, dI
\nonumber\\&&\times
( ({\rm cos}\alpha -1)_{AC'}\, -i{\bf R}^{a\alpha} \,{\bf n}^\alpha\,
(\tau^a )_{AC'}\,
{\rm sin}\,\alpha\,\,)\nonumber\\
&&\times
( ({\rm cos}\underline\alpha -1)_{BD'}\, -i{\bf R}^{b\beta}
\,\underline{\bf n}^\beta\,(\tau^b )_{AC'}\,
{\rm sin}\,\underline\alpha\,\,)\,\,,
\label{INE1}
\end{eqnarray}
where $dI$ is short for the instanton measure
\begin{eqnarray}
dI \equiv d^4z\,dn\,d{\bf R}\rightarrow n_0\,d^4z\,d{\bf R}\,\,.
\label{INE2}
\end{eqnarray}
The second equality holds for fixed instanton
density $n_0=1/\,{\rm fm}^{4}$ and size $\rho_0=1/3\,{\rm fm}$.
The tilde parameters follow from the untilded ones by setting $\theta=\pi/2$.
We note that $\underline{\tilde\gamma}=\underline\gamma=|\vec z|$. Note
that only the combination ${\bf R}\, {\bf R}$ survives after analytically
continuing to Minkowski space and taking the large $\sqrt{s}$  limit.

The one-instanton contribution to the parton-parton
scattering amplitude survives only in the color-changing channel a situation
reminiscent of one-gluon exchange~\cite{sz00}. As a result, the quasi-elastic
parton-parton cross section receives a finite two-instanton contribution at
large $\sqrt{s}$. The unitarized parton-parton partial differential
cross section reads
\begin{equation}
\frac{d\sigma}{dt} \approx \frac 1{s^2} \,\sum_{CD} \,\left| \,{\cal T}_{AC}^{BD}\,\right|^2\,\,,
\label{QQI5}
\end{equation}
with the averaging over the initial colors $A,B$ understood.
Simple algebra followed by the analytical continuation
$\theta\rightarrow -iy$, yield~\cite{sz00}
\be
\frac{d\sigma}{dt}\approx \frac{16\,n_0^2}{N_c^2(N_c^2-1)}\,
\left|\,\int\,db\,e^{iq\cdot b}\,F_{ss} \left(\frac b{\rho_0}\right)\right|^2\,\,.
\label{PARTIAL}
\ee
The one-instanton form factor $F_{ss}$ is defined as
\be
F_{ss}\left(\frac{b}{\rho_0}\right) =
\int\, d^4z\, \frac {(z_\perp -b)\cdot
z_\perp}{\tilde\gamma\,\underline{\tilde\gamma}}\,\,{\rm sin}\,\tilde\alpha\,
{\rm sin}\tilde{\underline\alpha}\,\,.
\label{FORMFACTOR}
\ee
In terms of (\ref{PARTIAL}), the quasi-elastic two-instanton contribution
to the forward parton-parton scattering amplitude is
\be
\sigma (t=0) \approx&&\frac {16\,n_0^2}{N_c^2 (N_c^2-1)}
\nonumber\\&&\times\int_0^\infty\,dq_\perp^2\,
\left|\,\int\,db\,e^{iq_\perp\cdot b}\,F_{ss} \left(\frac {b}{\rho_0}\right)\,\right|^2\,\,,
\label{LOW1}
\ee
which is finite at large $\sqrt{s}$. Hence, for forward
scattering partons in the instanton vacuum model, we expect~\cite{sz00}
\be
\sigma_{qq}\approx \pi\rho_0^2\,\,\kappa_0^2\,\,,
\ee
which is suppressed by two powers of the density. (\ref{LOW1})
is the instanton-induced generalization of the 2-gluon result derived 
by Low~\cite{LOW}.

\subsection{No Odderon}

In the early model  by Low~\cite{LOW} and Nussinov~\cite{NUSSINOV}, 
the near-forward high-energy
scattering amplitude is described by a perturbative two-gluon exchange
in the t-channel, which is C-parity even. Hence the $qq$ and 
$\bar q q$ cross sections are the same to this order, a result
that appears to be supported by experiment. Indeed, the difference 
$\sigma_{\bar pp}-\sigma_{pp}$ decreases at large
$\sqrt{s}$~\cite{odd_exp}.

However perturbation theory also allows for higher order corrections,
e.g. SU(3) allows for a colorless combination of 3
gluons. Perturbatively, the odderon/pomeron ratio is $O(\alpha_s)$
and not as suppressed as the data shows. To fix this problem, a number
of ideas have been put forward some of which relying on nucleon specifics 
(quark-diquark structure \cite{odd_th}) to cancel the odderon. If that
is the case, the odderon should still be observable in other hadronic
reactions.

In contrast, instanton-induced processes at high energy do not suffer
from the drawbacks of higher-order corrections. Indeed, even though
our quasi-elastic and even inelastic (see below) amplitudes
sum up an indefinite number of gluons, switching a quark to antiquark 
on the external line amounts to flipping the sign of the corresponding
${\rm sin}\,\alpha$ contribution. As there is no interference between
these and the ${\rm cos}\,\alpha$ terms at high energy, there is no
odderon in the instanton induced amplitudes. This is easily understood
by noting that an instanton is an SU(2) instead of an SU(3) field, for
which the fundamental (quark) and the adjoint (antiquark)
representations are equivalent.

\subsection{Weak Field Limit}
\label{sec_renorm_prop}

In the weak-field limit, most of our results~\cite{sz00} simplify with
interesting consequences on conventional perturbation theory. Indeed,
instanton-induced amplitudes involve integration over the
instanton (antiinstanton) center of mass $z$. So for fixed $z$ and large
impact parameter, the instanton field is weak~\footnote{The
shift $-{\bf 1}\rightarrow +{\bf 1}$ amounts to a change from regular
to singular gauge with no consequences for our analysis except in 
cancelling the identity in ${\bf W}$.}
\be
{\bf W} -{\bf 1}\approx -i n^a\,\tau^a\,\frac{\pi\rho_0^2}{2\gamma^2} \,\,,
\label{WE1}
\ee
which is conspicuous of a Coulomb field, familiar from perturbation
theory. We now discuss the consequences of this limit on quark-quark
scattering and gluon-gluon fusion to leading order.

\subsubsection{$QQ\rightarrow QQ$}

Inserting (\ref{WE1}) into the quark-quark scattering amplitude yields
after averaging over the global color orientations ${\bf R}$ to
\be
&&{\cal T}_{AB, CD} \approx 2is\,\kappa_0\,\frac{\pi^2}{{\rm
tan}\theta}\,(\tau^a)_{AC}(\tau^a)_{BD}\nonumber\\
&&\int db_{\perp}\,e^{iq_{\perp}b} 
\int dz_3\,dz_3'\,dz_{\perp} \,\,\frac{z_-\cdot z_+}
{(z_3^2+z_-^2)({z_3'}^2+z_+^2)}\,\,,
\label{WE2}
\ee
where we have defined $z_{\pm}=z_\perp\pm b/2$. The z-integrals
in (\ref{WE2}) diverge logarithmically. This divergence is similar
to the one encountered in perturbation theory~\cite{sz00} through
the exchange of a t-channel gluon at fixed impact parameter $b$, i.e.
\be
{\cal T} (\theta , b) = \frac {2\alpha_s}{{\rm
tan\,\theta}}\,{\rm ln}\left(\frac Tb\right)\,\,.
\label{WE3}
\ee
Hence (\ref{WE2}) can be interpreted as the instanton induced
renormalization of the perturbative gluon-exchange result 
(\ref{WE3}), with
\be
2\alpha_s\,\left(
1 + 2\pi^3 \,\frac{\kappa_0}{\alpha_s}\,\right)\,\,. 
\label{WE4}
\ee
The second contribution stems from the tail of the instanton 
in the weak field limit. It is natural to include this term
with the perturbative one-gluon exchange, subtracting it from
the truly instanton-induced amplitude. The latter is infrared
finite. A similar subtraction will also be needed in the inelastic
regime (see below).

\subsubsection{$gg\rightarrow g$}

In the weak-field limit the fusion $gg\rightarrow g$ is best
analyzed in momentum space using the Fourier transform of 
(\ref{WE1}),
\be
A_\mu^a (k) = \frac{\pi\rho_0^2}{gk^2}\,{\bf R}^{ab}\,\eta^a_{\mu\nu}\,k_\nu\,\,.
\label{WE5x}
\ee
In terms of (\ref{WE5x}) the fusion reaction in a single
instanton follows from 
\be
&&\Gamma^{a1,a2,a3}_{\mu1,\mu2,\mu3} = \,\left(\frac{\pi\rho^2_0}{g}\right)^3\,
({\bf R}^{a1,b1}\,\bar \eta_{b1,\mu1,\nu1}
k1_{\nu1})\nonumber\\
&& ({\bf R}^{a2,b2}\bar \eta_{b2,\mu2,\nu2} k2_{\nu2})
({\bf R}^{a3,b3}\bar \eta_{b3,\mu3,\nu3} k3_{\nu3}) \,\,,
\label{WE5}
\ee
after using the LSZ amputated form of (\ref{WE5x}).
Since ${\bf R}$ is isomorph to the (3,3) representation
of SU(2), we note the identity
\be
{\bf R}^{ab}\,{\bf R}^{cd}\,{\bf R}^{ef} = &&\frac 16 
\epsilon^{ace}\,\epsilon^{bdf}\nonumber\\
&& \bigoplus_{j=1}^{3} \,(2j+1, 2j+1) \,\,,
\label{WE6}
\ee
in terms of irreducible representations. For convenience,
only the (1,1) contribution is explicitly quoted. Using
(\ref{WE6}) and the identities for the 't Hooft symbol,
we obtain
\be
&&\Gamma^{a1,a2,a3}_{\mu1,\mu2,\mu3} (k1, k2, k3)
 =\, \left(\frac{\pi\rho^2_0}{g}\right)^3\,
\epsilon^{a1,a2,a3}\\
&&\times (k1\cdot k3\,\left( \delta_{\mu2,\mu3} \, k2_{\mu1} -
\delta_{\mu1, \mu2} \, k2_{\mu3}\right)\nonumber\\
&&\qquad +k1\cdot k2\,\left( \delta_{\mu1,\mu3} \, k3_{\mu2} -
\delta_{\mu2, \mu3} \, k3_{\mu1}\right)\nonumber\\
&&\qquad +k2\cdot k3\,\left( \delta_{\mu1,\mu2} \, k1_{\mu3} -
\delta_{\mu1, \mu3} \, k1_{\mu2}\right)\nonumber\\
&&\qquad +k1_{\mu2}\,k2_{\mu3}\,k3_{\mu1} 
-k1_{\mu3}\,k2_{\mu1}\,k3_{\mu2}\nonumber\\
&&\qquad -\epsilon_{\mu1,\alpha, \beta,\mu3}\,k1_{\alpha}\,k2_\beta\,k3_{\mu2} -
\epsilon_{\mu1,\alpha,\mu2,\beta}\,k1_\alpha\,k2_{\mu3}\,k3_\beta\nonumber\\
&&\qquad -k2\cdot k3\,\epsilon_{\mu1,\mu2,\mu3,\alpha} \,k1_\alpha
+\delta_{\mu2,\mu3}\,\epsilon_{\mu1,\alpha,\beta,\gamma}\,
k1_{\alpha}\,k2_{\beta}\,k3_{\gamma})\nonumber
\ee
where only the $(1,1)$ contribution was retained.
Since $k1+k2+k3=0$ only part of this expression matches
kinematically the standard perturbative 3-gluon vertex,
thereby producing an instanton-induced contribution of
relative strength 
$\kappa_0/\alpha_s^2$. It can be regarded as the
instanton-induced contribution to the elementary 
BFKL ladder. We note that the induced strength in
the gluon fusion is stronger than the relative strength 
of $\kappa_0/\alpha_s$ seen in the exchange (\ref{WE4}). 
This is a general feature of the inelastic processes as we now
discuss.

\section{Inelastic Scattering}

To address inelastic amplitudes with instantons, the eikonal
approximation has to be relaxed. To achieve that and elucidate
further the character of the s-channel kinematics, we first derive
a general result for on-shell quark propagation in a localized 
background field in Minkowski space. We then show how this result 
can be applied to instanton dynamics to analyze inelastic 
parton-parton scattering at high energy beyond the eikonal 
approximation and ladder graphs. For simplicity, all the
instanton algebra will be carried out explicitly for $N_c=2$.

\subsection{Beyond the Eikonal Approximation}

An on-shell massless quark propagating through a localized
background $A(x)$ with initial and final momenta $p_1$ and 
$p_2$ follows from LSZ reduction
\be
{\cal S} (p_1, p_2; A (x)) \equiv 
<p_2 | i\vec{\rlap/\partial}\,{\bf S}_F (x)\, i\vec{\rlap/\partial} |p_1>\,\,,
\label{BE1}
\ee
with $i\rlap/\nabla\,{\bf S}_F =-1$ the background (Feynman) propagator
in the instanton field. At large $p_+$ momentum, the quark propagates
on a straight-line along the light-cone. This limit can be used to
organize (\ref{BE1}) in powers of $1/p_+$. The result is
\be
&&{\cal S} (p_2, p_1; A(x) ) =e^{i(p_2-p_1)x}\,
\nonumber\\&&\times\,\overline{u} (p_2) \, g\rlap/A\,
\sum_{n=0}^{\infty}
\left(\frac i{2p_1\cdot\nabla}\,\rlap/\nabla\,\frac{\rlap/p_1 \,\gamma_0}{2p_{10}} \,
\rlap/\nabla\right)^n \nonumber\\
&&\times\,
{\bf W}_{-} (x_{1+}, x_{1-}, x_\perp )\, u(p_1) \,\,,
\label{BE2}
\ee
where
\be
&&{\bf W}_{-} (x_{1+}, x_{1-}, x_\perp ) = \nonumber\\
&&{\bf P}_c \,{\rm exp} \left(-\frac {ig}2\int_{-\infty}^{x_{1+}}
dx_+' \,A_- (x_+', x_{1-}, x_\perp )\right)\,\,.
\label{BE3}
\ee
The line-integral is carried along the $p_1$-direction of the original
quark line with $x_{1\pm}= (p_{0}x_0 \pm \vec{p}\cdot\vec{x})/p_0$. 
In the limit $p_{10}\rightarrow\infty$, only the $n=0$ term contributes
with $x_{1\pm}=x_{\pm}$ being just the light-cone coordinates, thereby
reproducing the eikonal result (\ref{BE3}). The higher-order terms are 
corrections to the eikonal-result, with the $n=1$ term accounting for 
both recoil and spin effects.

\subsection{Inelastic Amplitude}

The imaginary part of the quark-quark inelastic amplitude
follows from unitarity. Schematically, 
\be
{\rm Im} \, {\cal T}_{if} = {\cal T}_{in} \,\sigma_{nn} \, {\cal T}_{nf}^*\,\,,
\label{FI1}
\ee
where $\sigma_{nn}$ accounts for the phase space of the 
propagating quarks and emitted intermediate gluons. The total
cross section follows then from the optical theorem 
$\sigma= {\rm Im} \, {\cal T}/4s$. Using the result (\ref{BE2}),
we have for the total cross section in Minkowski space
\be
\sigma = &&\frac 1{4s} \,{\rm Im}\, \sum_{CD} \,\int
d[A]\,d[{A'}]\,e^{i(S[A]-S[{A'}] )}\nonumber\\
&&\times\int \frac{d^3K_1}{(2\pi )^3}\,\frac{d^3K_2}{(2\pi )^3}
\frac 1{2K_{10}}\frac 1{2K_{20}}\nonumber\\
&&\times\left( \sum_{n=0}^{\infty} \frac 1{n!}\prod_{i=1}^n
\int \frac {d^3k_i}{(2\pi )^3}\, \frac 1{2k_{i0}} \, {\cal A} (k_i)
\,{\cal  A'}^{*} (k_i) \right)\nonumber\\
&&\times\frac 1{VT} \int dx\,dy\,dx'\,dy'\,\nonumber\\
&&\times e^{i(K_1-p_1)x +(K_2-p_2)y - (K_1-p_1)x' -i(K_2-p_2)y'}\nonumber\\
&&\times{\cal S}_{AC} (K_1, p_1; A (x))\,
{\cal S}^*_{AC} (K_1, p_1; A' (x'))\,\nonumber\\&&\times
{\cal S}_{BD} (K_2, p_2; A (y))\,
{\cal S}^*_{BD} (K_2, p_2; A' (y'))\,\,.
\label{FI2}
\ee
The functional integration is understood over gauge-fields (to be
saturated by instantons in Euclidean space after proper analytical
continuation), with ${\cal A} (k)$ (${\cal A}' (k)$) the Fourier 
transform of the pertinent asymptotic of $A$ ($A'$) evaluated on 
mass shell. Similar expressions 
were used for sphaleron-mediated gluon fusion~\cite{weakinst}. The
difference with the present case is the occurrence of quarks in both
the initial, intermediate and final states. The sum in (\ref{FI2}) 
exponentiates into the so-called R-term, which acts as an induced interaction
between the $A$ and $A'$ configurations in the double functional
integral (\ref{FI2}). We will refer to it as $S(A, A')$.

The gauge fields carried inside the on-shell quark propagators
${\cal S}$ involve virtual exchange of background quanta with
no contribution to the cut. In contrast, the on-shell gluons
${\cal A} (k)$ are real and the sole contributors to the cut. 
The $n=0$ term in (\ref{FI2}) in the large $p_+$ limit reduces
to the quasi-inelastic contribution discussed above. The term
of order $n$ involves $n$-intermediate on-shell gluons plus
two on-shell quarks, and contributes to the bulk of the inelastic
amplitude.

The general result (\ref{FI2}) involves no kinematical approximation
regarding the in/out quark states. At high-energy, all $1/p_+$ effects
in (\ref{BE2}) can be dropped to leading ${\rm ln}\, s$ accuracy except
in the exponent. As a result, (\ref{FI2}) simplifies dramatically,
\be
&&\sigma \approx \frac 1{4VT} \,{\rm Im}\,\sum_{CD}\,\frac{1}{(2\pi)^6}
\,\int dq_{1+}\,dq_{1\perp}\,dq_{2-}\,dq_{2\perp}\nonumber\\
&&\times \int [dA][dA']\,e^{iS(A)-iS(A')+
iS(A,A')}\,\nonumber\\
&&\times\int dx_-dx_\perp dy_+dy_\perp \,
e^{\frac i2 q_{1+}x_--iq_{1\perp}
x_\perp +\frac i2 q_{2-}y_+ -iq_{2\perp}y_\perp}\nonumber\\
&&\times\left( {\bf W}_- (\infty, x_-, x_\perp) -{\bf 1}\right)_{AC}
\left( {\bf W}_+ (y_+, \infty, y_\perp) -{\bf 1}\right)_{BD}\nonumber\\
&&\times\int dx'_-dx'_\perp dy'_+dy'_\perp \,e^{\frac i2 q_{1+}x'_--iq_{1\perp}
x'_\perp +\frac i2 q_{2-}y'_+ -iq_{2\perp}y'_\perp}\nonumber\\
&&\times\left( {\bf W}_- (\infty, x'_-, x'_\perp) -{\bf 1}\right)^*_{AC}
\left( {\bf W}_+ (y'_+, \infty, y'_\perp) -{\bf 1}\right)^*_{BD}\,\,.
\nonumber\\
\label{FI3}
\ee
Overall, the scattering amplitude follows from the imaginary
part of a retarded 4-point correlation function  in Minkowski
space. This correlation function follows from a doubling of the
fields, a situation reminiscent of thermo-field dynamics.

To proceed further, some dynamical approximations are needed. 
Let us assume that the double-functional integral in (\ref{FI3})
involves some background field configurations characterized
by a set of collective variables (still in Minkowski space),
say $I=Z,{\bf R},\rho$, for position,
color orientation respectively. Let $z=Z-Z'$ be the relative
collective position. Simple shifts of integrations, produce
\be
e^{iQz} =e^{\frac i2 q_{1+}z_- +\frac i2
q_{2_-}z_+-i(q_1+q_2)_\perp\,z_\perp}\,\,,
\label{FI4}
\ee
with no dependence on $Z,Z'$ in the ${\bf W}$'s. The integration
over the location of the CM $(Z+Z')/2$ produces $VT$ which cancels the $1/VT$ in front,
due to overall translational invariance. The integration over the
relative coordinate $z$
produces a function of the invariant $Q^2=q_{1+}q_{2-}
-(q_1+q_2)_\perp^2$ because of Lorentz invariance. With this 
observation and to leading logarithm accuracy, we may rewrite
(\ref{FI3}) as follows
\be
&&\sigma \approx \frac 14  \,\,{\rm ln s}\,\,{\rm Im}\,\sum_{CD} 
\,\frac{1}{(2\pi)^6}\,\int dQ^2 \, dq_{1\perp} \, dq_{2\perp }\,\nonumber\\
&&\times\int\,dz\,d\dot{I}\,d\dot{I}'\,e^{iQz + iS(\dot{I})-
iS(\dot{I}')+ iS(\dot{I}, \dot{I}',z)}\nonumber\\
&&\times\int dx_-dx_\perp dy_+dy_\perp \,e^{-iq_{1\perp}
x_\perp -iq_{2\perp}y_\perp}\nonumber\\
&&\times\left( {\bf W}_- (\infty, x_-, x_\perp) -{\bf 1}\right)_{AC}
\left( {\bf W}_+ (y_+, \infty, y_\perp) -{\bf 1}\right)_{BD}\nonumber\\
&&\times\int dx'_-dx'_\perp dy'_+dy'_\perp \,e^{-iq_{1\perp}
x'_\perp -iq_{2\perp}y'_\perp}\nonumber\\
&&\times\left( {\bf W}_- (\infty, x'_-, x'_\perp) -{\bf 1}\right)^*_{AC}
\left( {\bf W}_+ (y'_+, \infty, y'_\perp) -{\bf 1}\right)^*_{BD}\,\,.
\nonumber\\
\label{FI33}
\ee
Note that the omitted exponents $e^{\frac i2 q_+x_-}$ etc. are
sub-leading in leading logarithm accuracy.
The dotted integrations no longer involve the collective variables
$Z,Z'$. The only left dependence on the relative variable
$z$ resides in the induced R-term. The appearance of ${\rm ln}\,s$
underlines the fact that the integrand in (\ref{FI33}) involves only 
$Q^2$ which is the transferred mass in the inelastic half of the
forward amplitude, and $q_{1,2\perp}$ which are the transferred
momenta through the quark form-factors.

All kinematical approximations in this section were carried
out in Minkowski space, a point stressed in our earlier work
\cite{sz00}. The result is (\ref{FI33}) to leading logarithm 
accuracy. This is one of our main result, showing that the
inelastic contributions to the forward quark-quark scattering
amplitude cause the latter to rise with ${\rm ln} s$, irrespective
of the background field used. The outcome (\ref{FI33}) is now ripe
for an analysis in Euclidean space using lattice Monte-Carlo
simulations or instantons as we now discuss.

\subsection{Instanton-Antiinstanton Interaction}

The general ${\cal W}$ correlation function made of the 
four ${\bf W}$'s in 
(\ref{FI3}) for fixed kinematics $Q, q_{1\perp}, q_{2\perp}$,
is best analyzed in Euclidean space, where the ${\bf W}$'s
are defined at a rapidity $\theta$ and $Q_E^2 = -Q^2 <0$.
The dominant background configurations at  $Q_E\,z\gg 1$ 
are instanton-antiinstanton configurations.  Specifically
\be
{\cal W} \approx &&\frac 14\, {n_0^2} \,
\sum_{CD} \int d^4z \, d{\bf R}\, d{\bf R'}
\,\,e^{iQ_E z} \,e^{-{\bf S} (z, {\bf RR'}^{-1} )}\nonumber\\
&&\times\,\int d^3x \, d^3y\, d^3x'\, d^3y' 
e^{-iq_{1\perp} (x -x') -iq_{2\perp} (y-y') }\nonumber\\
&&\times
( ({\rm cos}\alpha -1)_{AC}\, -i{\bf R}^{a\alpha}
\,{\bf n}^\alpha\,(\tau^a )_{AC}\,
{\rm sin}\,\alpha\,\,) \nonumber\\
&&\times
( ({\rm cos}\underline\alpha -1)_{BD}\, -i{\bf R}^{b\beta}
\,\underline{\bf n}^\beta\,(\tau^b )_{BD}\,
{\rm sin}\,\underline\alpha\,\,)\nonumber\\
&&\times
( ({\rm cos}\alpha' -1)_{AC}\, +i{\bf R'}^{a'\alpha'} \,{\bf n'}^{\alpha'}\,
{(\tau^{a'} )^*}_{AC}\,
{\rm sin}\,\alpha'\,\,)\nonumber\\
&&\times
( ({\rm cos}\underline{\alpha'} -1)_{BD}\, +i{\bf R'}^{b'\beta'}
\,\underline{\bf n'}^{\beta'}\,\tau^{b'\,*}_{BD}\,
{\rm sin}\,\underline{\alpha'}\,\,)\,\,,\nonumber\\
\label{FI44}
\ee
where the variables $x,x'$ are defined on a tilted Wilson line
of angle $\theta$ with $x_4$, and $y,y'$ on an untilted Wilson 
line running along $y_4$. The instanton-antiinstanton interaction
is known precisely in leading order,
\be
{\bf S} (z, {\bf RR'}^{-1})= \frac {4\pi}{\alpha_s} \, (3 \,u_0^2-1)\,
\left( -2 \frac {\rho_0^2}{z^4} + 8 \frac {\rho_0^6}{z^6} + ...\right) \,\,,
\label{FI5}
\ee
with $u_0={\rm Tr} U/2$  and the unitary
parameterization of the orthogonal matrices
\be
\left({\bf RR'}^{-1}\right)^{\alpha\beta} = 
\frac 12 \, {\rm Tr} ( U\tau^\alpha\,U^\dagger\tau^\beta )\,\,.
\label{FI6}
\ee
The first contribution in (\ref{FI5}) is the well-known dipole
contribution, which is known to match exactly the R-contribution
stemming from the exponentiation of retarded gluons from instanton
vertices~\cite{weakinst}.

There are many contributions in (\ref{FI44}). However, we note that
in Minkowski space, the dominant contribution while continuing in
$\theta$ involves ${\bf RRRR}$. From here on, it will be the only
one retained. With this in mind, we carry first the integration over
the collective variable ${\bf R}$ and ${\bf R'}$ in the SU(2) case
by explicitly carrying part of the group integration. Setting 
$u_0={\rm cos}\chi$, and averaging over SU(2) gives
\be
&&{\bf RRRR} = \\
&&+\frac 2\pi \, I_1\, {\bf n}\cdot{\bf n'} \,
\underline{\bf n}\cdot\underline{\bf n'}\nonumber\\
&&+\frac {2}{3\pi}\, I_2\,
\left(-2{\bf n}\cdot{\bf n'} \,\underline{\bf n}\cdot\underline{\bf n'}
+ 4 ( {\bf n}\cdot\underline{\bf n}\,{\bf n'}\cdot\underline{\bf n'}
     -{\bf n}\cdot\underline{\bf n'}\,{\bf n'}\cdot\underline{\bf n} )\right)
\nonumber\\
&&+\frac {2}{15\pi}\, I_3\,
\left(-{\bf n}\cdot{\bf n'} \,\underline{\bf n}\cdot\underline{\bf n'}
+ 4 ( {\bf n}\cdot\underline{\bf n}\,{\bf n'}\cdot\underline{\bf n'}
     +{\bf n}\cdot\underline{\bf n'}\,{\bf n'}\cdot\underline{\bf n}
)\right)\,\,,\nonumber
\ee
with 
\be
I_k = \int_0^\pi\,d\chi\, {\rm sin}^{2k} \chi\,\,{\rm
cos}^{6-2k}\chi\,\,e^{-{\bf S} (z, {\rm cos}^2\chi )}
\label{FI8}
\ee
for $k=1,2,3$. Inserting (\ref{FI8}) back into (\ref{FI44}) and
performing the analytical continuation back to Minkowski space
shows that only the combination 
\be
\left({\bf n}\cdot\underline{\bf n}\,{\bf n'}\cdot\underline{\bf n'}
     +{\bf n}\cdot\underline{\bf n'}\,{\bf n'}\cdot\underline{\bf n}\right)\,\,,
\nonumber
\ee
survives. The result is
\be
&&{\cal W} (Q, q_{1\perp}, q_{2\perp} )
 = (8\pi^5)^\frac 12\,\,\,{\bf K} (q_{1\perp} , q_{2\perp})\nonumber\\
&&\times{\rm Im}\,\, n_0^2\,\int_0^\infty dR \, \left(\frac RQ \right)^{\frac 32}\, 
\int_0^\pi\,d\chi\,{\rm sin}^6\chi\,e^{QR-{\bf S}(R,{\rm cos}^2\chi)}\,\,,
\nonumber\\
\label{FI9}
\ee
with the induced kernel
\be
{\bf K} (q_{1\perp} , q_{2\perp}) =
|{\bf J} (q_{1\perp}) \cdot {\bf J} (q_{2\perp})
+ {\bf J} (q_{1\perp})\times {\bf J} (q_{2\perp}) |^2
\,\,,
\label{FI10x}
\ee
We have introduced our generic instanton-induced  form-factor
\be
{\bf J} (q_{\perp}) = \int dx_3 \,dx_\perp \, e^{-iq_\perp x} \,
\frac{x_\perp}{|x|}\,{\rm sin} \left( \frac {\pi\,
|x|}{\sqrt{x^2+\rho_0^2}}\right)\,\,.
\label{FI10}
\ee
which is purely imaginary,
\be
{\bf J} (q_{\perp}) =&& -i \frac{\hat{q}_\perp}{\sqrt{q_\perp}}
\,\,\int_0^\infty\,dx\,J_{3/2} (q_\perp x)\,\nonumber\\
&&\times\,\left( (2\,\pi x)^{3/2}\,{\rm sin}\left(\frac {\pi\,
|x|}{\sqrt{x^2+\rho_0^2}}\right)\right)\,\,.
\label{FI10xx}
\ee
Here $J_{3/2}$ is a half-integer Bessel function. In the weak-field
limit the instanton contributes a term ${}^3\sqrt{x}/x^2\approx 1/\sqrt{x}$ 
that causes the instanton-induced form factor to diverge. This
divergence is analogous to the one encountered in $QQ\rightarrow QQ$.
The behavior of (\ref{FI10xx})  is shown in Fig.~\ref{fig_J} (top points).
Apart from the unphysical (perturbative) singularity at small $q_\perp$,
the instanton-induced form factor can be parameterized by a simple
exponential
\be
\label{J_param}
{\bf J} (q_\perp) \approx -i\,{\hat{q}}_\perp\,50\, e^{-1.3\, q_\perp\, \rho_0} \,\,,
\ee
which is the solid line shown in Fig.~\ref{fig_J}. We note that this
is very different from just the Fourier transform of the instanton
field used as a form-factor in~\cite{KKL} (see further discussion
in section VIB below). Throughout, the tail of the instanton will 
be subtracted resulting into a renormalization of the perturbative result.
\begin{figure}[h]
\vskip 0.2in
\includegraphics[width=2.2in]{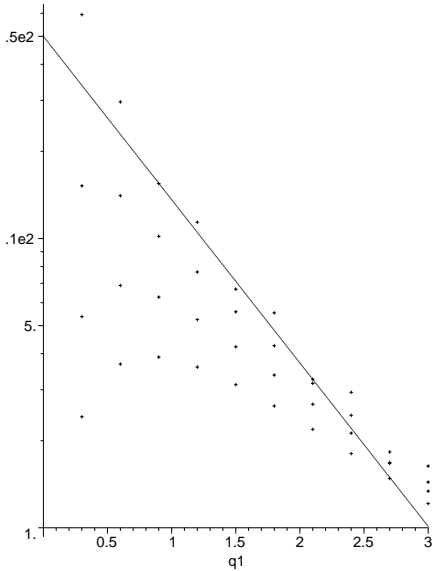}
\vskip 0.2in
\caption[]{
 \label{fig_J}
The induced instanton form-factor $|{\bf J}(q_\perp)|$ (points) and its
parameterization (\ref{J_param}) (solid line) versus
$q1=q_\perp \rho_0$. The four set of points
(counting from top to bottom) correspond to different parameterizations
of the instanton shape (see section VC below)   
${\bf a}=0,0.25,0.50,0.75$ (top to bottom).}
\end{figure}

The imaginary part of (\ref{FI10}) is readily assessed in the
dipole approximation by retaining only the first contribution in
(\ref{FI5}). Carrying the $R$ and $\gamma$ integrations by saddle 
point we obtain a purely imaginary result to leading order owing to
the unstable mode around the instanton-antiinstanton configuration.
The result for the total cross section is
\be
\sigma \approx &&\pi\rho_0^2\,\kappa_0^2\,{\rm ln
\,s}\,\frac {16}{15} \,\frac 1{(2\pi)^8}\,\int\,dq_{1\perp}\,dq_{2\perp}\,
\,{\bf K} (q_{1\perp} , q_{2\perp})\nonumber\\
&&\times{\rm Im}\,i\,\,
{\bf C}\,\,\alpha^{\frac 3{10}} \,
\int_{(q_1+q_2)_{\perp}^2}^\infty
 dQ^2\, \frac{e^{\frac 52 (2\pi Q^4/\alpha)^{1/5}}}{Q^{37/10}}\,\,,
\label{FI11}
\ee
with ${\bf C}$ a number inherited from the R- and $\chi$-saddle points, 
\be
{\bf C} = 10\,\sqrt{6}\,(2\pi)^{\frac
15}\,\left(\frac{16\pi}{3}\right)^3\,\,.
\ee
All the integrations are over { dimensionless} variables re-expressed 
in units of the instanton size $\rho_0$. From here on, this will be
assumed unless indicated otherwise. The Q-integration diverges 
at the upper end. This is not surprising since
the dipole-approximation is valid for small invariant mass $Q^2$ or
large separation $z$. As $Q^2$ increases, the higher order contributions
in (\ref{FI5}) become important.
This is the realm of the streamline as we now discuss.

\subsection{Streamline}

In the Euclidean  regime $Q_Ez\approx 1$, the instanton-antiinstanton 
overlaps as their interaction becomes strong. In this case, it is more
appropriate to use the streamline configuration, which is a gauge
configuration that interpolates between an instanton-antiinstanton 
asymptotically and the vacuum, following the path of least action
(valley). A very good parameterization of the streamline
follows from conformal symmetry. In particular
\be
{\bf S} (z, {\rm cos}^2\chi ) = a(z) + b(z)\, {\rm cos}^2\,\chi +
c(z) \, {\rm cos}^4\,\chi\,\,,
\label{FI12}
\ee
where $a(z)$, $b (z)$ and $c (z)$ are known functions of
$z$~\cite{streamline}.  Using (\ref{FI12})
in the saddle point approximation carried above, allows for a 
better assessment of the $Q^2$-integrand in (\ref{FI11}). In
particular,
\be
{\bf C}\,\alpha^{\frac 3{10}} \,
\frac{e^{\frac 52 (2\pi Q^4/\alpha)^{1/5}}}{Q^{37/10}}
\rightarrow B(Q) = e^{F(Q/Q_s)}\,\,,
\label{FI13}
\ee
where $Q_s^2$ is the sphaleron invariant mass squared. The streamline
configuration allows us to extend the validity of the dipole
approximation to higher $Q^2$, through the holy-grail function $F$. 
The specific form of $F$ will not be needed if unitarization takes
place as we now show.

\subsection{Multi-instantons and Unitarization}

The inelastic contributions to the quark-quark scattering
amplitude causes the total cross section to grow rapidly
with the longitudinal energy transferred $Q^2$. Since the
streamline configuration leads eventually to the vacuum, 
one may be tempted to argue that this generates
unsuppressed multi-gluon production with unbounded
cross section~\cite{ZAKHAROV}. This conclusion is physically
incorrect.

Indeed, in the analogous problem of baryon number violation
in the standard model, Zakharov~\cite{ZAKHAROV} has argued that for $
Q\approx Q_s$
the rise in the cross section has to stop because of unitarity 
constraints. Maggiore and Shifman~\cite{MS} suggested that as
$Q^2$ increases, or equivalently as the instanton-antiinstanton
separation decreases, multi-instanton effects become important.
Unitarization can be simply enforced by resumming a chain of
alternating instanton-antiinstanton configurations, leading
to a unitarized amplitude confirming Zakharov's observations.

Following on Maggiore and Shifman suggestion in the baryon
number violation problem, we perform the large $Q^2$ 
integration in (\ref{FI11})  using iterated multi-instanton contributions.
This is similar to (although different from) the usual treatment of
resonances, when the attractive interaction is iterated and
 leads to a Breit-Wigner result.  Specifically,
the imaginary part in (\ref{FI11}) now reads
\be
\sum_{n=1}^\infty\, \kappa_0\,
\left( \kappa_0\,i\,B(Q)\right)^n\,\,,
\label{FI14}
\ee
for alternating insertions of instantons and antiinstantons (chain).
Each factor of $iB$ results from the insertion of an extra instanton
or antiinstanton on the chain, producing a bond with an extra unstable
mode. Hence, the total cross section is
\be
\sigma \approx &&\pi\,\rho_0^2\,\,{\rm ln \,s}\,
\frac {16}{15}\,\frac 1{(2\pi)^8}\,\int\,dq_{1\perp}\,dq_{2\perp}\,
\,{\bf K} (q_{1\perp} , q_{2\perp})
\nonumber\\&&\times
\kappa_0\,\,
\int_{(q_1+q_2)_{\perp}^2}^\infty
 dQ^2\, \frac{\kappa_0\,B(Q)}{1+ \kappa_0^2\,B(Q)^2}\,\,.
\label{FI15}
\ee
The integrand in (\ref{FI15}) rises with $B(Q)$ as expected in the
small $Q$ regime and falls off as $1/B(Q)$ due to unitarization. The
dominant contribution takes place at the sphaleron invariant mass
\be
B(Q_s) \approx 1/\kappa_0
\label{FI16}
\ee
for which the total cross section (\ref{FI5}) becomes
\be
\sigma \approx &&\pi\rho_0^2\,\,{\rm ln \,s}\,\,
\kappa_0\,
\frac {16}{15}\,\frac 1{(2\pi)^8}\,
\int\,dq_{1\perp}\,dq_{2\perp}\,
\,{\bf K} (q_{1\perp} , q_{2\perp})\,\,.
\label{FI17}
\ee
Note that under the condition (\ref{FI16}) the $Q^2$-integration
amounts to a number of order 1, a measure of the area under the curve
peaked at $Q_s$ with maximum $1/2$ and width
of order 1. The rise in the partial inelastic cross section due to
multi-instanton effects results into an increase of the
cross section by one power of the diluteness factor
which is about a 100-fold increase. The inelastic cross 
section grows logarithmically with $s$, in contrast to our
original quasi-elastic estimate~\cite{sz00}. 

The energy following from (\ref{FI16}) implies
that half of the original instanton-antiinstanton action
of $2\times(8\pi^2/g^2)$ is compensated by their attraction. In 
other words, the s-exchange in the inelastic process starting
from the vacuum is {\it half-instanton}. This is the transition
from vacuum to a static QCD sphaleron. This leads us to the 
following important observation: at high energy, the inelastic
s-channel contributions to parton-parton scattering in leading 
logarithm approximation are QCD sphalerons.

\section{Soft Pomeron from Instantons}

In this section we will show that in the semiclassical analysis
the parton-parton cross-section increases at most as 
${\rm ln}\,{s}$, while the hadron-hadron cross section as
a polynomial in ${\rm ln}\,s$, with a degree fixed by the number
of hard collisions in the transverse plane. We show that Reggeization
follows when the number of collisions becomes large.

\subsection{Intercept}

The parton-parton cross section at large $\sqrt{s}$ with $-t/s$ small, 
receives contributions from both perturbative gluons and instantons.
Generically, the perturbative contributions are
\be
\sigma_* (s,t) \approx \pi\,\rho_*^2\,
\left( ({\alpha_s}/{\pi})^2 + \# \,({\alpha_s/\pi})^3
\,{\rm ln}\, s + ...\right)\,\,,
\label{POM0}
\ee
where $1/\rho_*$ is some perturbative QCD cutoff scale, at which the 
running ${\alpha_s}$ should be defined. Instantons  contribute partly to
the constant part of the cross section term ~\cite{sz00},  along with other
non-perturbative contributions. Fortunately, we do not have to
go into all of this by noting that the magnitude of the total quark-quark 
cross section can be assessed in a model independent way. Indeed, for 
$\sqrt{s}\approx 30\,{\rm GeV}$ the $pp$ cross section does not
grow yet, so that the sea quarks and gluonic contributions can be
ignored. Hence, a simple {additive quark model} estimate yields
\be
\sigma_{qq}= {1\over 9}\, \sigma_{pp}\approx 3.3\,{\rm  mb}\,\,
\label{ESTIM}
\ee
where the inelastic cross section $\sigma_{pp}\approx 30$ mb has been used.
Setting $\sigma_{qq}=\pi r_0^2$ we find that (\ref{ESTIM}) reflects on a
typical scattering disk of radius $r_0\approx 1/3 \,{\rm fm}$.  

The instanton contribution to the inelastic process, yields a
logarithmically growing cross section
\be
\sigma (s, t) \approx 
\pi\,\rho_0^2\, 
\left( \#\, \kappa_0\, {\rm ln}\, s + ...\right) \,\,.
\label{POM1}
\ee
Hence,
\be
\sigma (s,t) \approx  \pi\,\rho_*^2\, 
\,({\alpha_s}/{\pi})^2 +\pi\,\rho_0^2\, 
\,\Delta (t)\, {\rm ln}\, s  
\label{POM2}
\ee
with 
\be
\Delta (0) = \kappa_0\,\frac {16}{15}\,\frac 1{(2\pi)^8}\,
\int\,dq_{1\perp}\,dq_{2\perp}\,
\,{\bf K} (q_{1\perp} , q_{2\perp})\,\,.
\label{POM3}
\ee
Using (\ref{FI10x}) we note that the spin-0 and spin-1 parts
contribute equally to the intercept, giving
\be
\Delta (0) = \kappa_0\,
\frac {16}{15}\,\frac 1{(2\pi)^8}\,
\left(\int_0^\infty\,dq\,{\bf G}^2 (q)\right)^2\,\,,
\label{POM3x}
\ee
where we have defined the scalar form-factor ${\bf G}$ as
\be
{\bf J} (q) = -i\frac{\hat{q}}{\sqrt{q}}\,\frac{{\bf G} (q)}{\sqrt{2\pi}}\,\,.
\label{POM3xx}
\ee
A numerical estimate of (\ref{POM3x}) can be made using the
parameterization (\ref{J_param}) which removes the unphysical
singularity at $q_\perp =0$. The result is
\be
\Delta (0)= 2.37\,\kappa_0\approx 0.03\,\,, 
\label{MOD3}
\ee
which is smaller than the phenomenological intercept of 0.08~\cite{DL}
for the soft pomeron. This is of no concern, since the instanton
density itself is known within a factor 2 and maybe updated upward
\footnote{Lattice studies confirm the ``standard'' value quoted above,
in the ``deep cooling'' regime, where only well separated instantons
relevant for chiral symmetry breaking are retained. The vacuum
contains also a large density of  close instanton-antiinstanton
pairs eliminated by ``cooling''. Those are
irrelevant for the light quark condensate
 but are relevant for high energy scattering.}.
Additional effects absent in the present instanton
estimate will be discussed below. Our main conclusion is that
the smallness of the soft pomeron intercept directly reflects on the 
diluteness of the instantons in the QCD vacuum, thereby providing us
with a first hand empirical glimpse to this important parameter.

\subsection{Slope}

The t-dependence in the parton-parton cross section follows
from the inelastic processes with net momentum flow in the
t-channel. We can change minimally the forward scattering
amplitude to allow for this, leading to the following expression 

\be
\Delta (t) = \kappa_0\,\frac{16}{15}\,\frac 1{(2\pi)^8}\,
\int\,dq_{1\perp}\,dq_{2\perp}\,
\,{\bf H} (q_{1\perp} , q_{2\perp}; t)\,\,,
\label{POMA}
\ee
with the new t-dependent induced kernel ($t=-q_\perp^2$)
\be
&&{\bf H} (q_{1\perp} , q_{2\perp}; t) \equiv \nonumber\\
&&( {\bf J} (q_{1\perp} - q_\perp/2 ) 
\cdot {\bf J} (q_{2\perp} -q_\perp/2)\nonumber\\
&&+ {\bf J} (q_{1\perp}-q_\perp/2 )\times {\bf J} (q_{2\perp}-q_\perp/2)
)\nonumber\\
&&\times( {\bf J} (q_{1\perp} + q_\perp/2 ) 
\cdot {\bf J} (q_{2\perp} + q_\perp/2)\nonumber\\
&&+ {\bf J} (q_{1\perp}+q_\perp/2 )\times {\bf J} (q_{2\perp}+q_\perp/2)
)^*\,\,.
\label{POMB}
\ee 
The form factors are defined as in (\ref{FI10}-\ref{FI10xx}).
For $t\approx 0$, we have
$\Delta (t) \approx \Delta (0) + t\, \Delta' (0$. The slope
parameter $\Delta' (0)$ follows from a Taylor expansion of
(\ref{POMA}) {\it after} integration. Using (\ref{J_param})
which removes the unphysical singularity at $q_\perp\approx 0$
(related to the perturbative singularity discussed earlier
in $QQ\rightarrow QQ$ scattering), we can perform the double integrations in
(\ref{POMA}) to obtain ${\bf H} (q_\perp^2)$ as shown in 
Fig.~\ref{fig_H}. Modulo the pre-factors in (\ref{POMA}) this
is just the pomeron trajectory for $t<0$ (physical region).
The upper curve refers to instantons with unmodified vacuum
sizes (${\bf a}=0$), while the lower curve corresponds to 
slightly smaller instantons (${\bf a}=0.25$).
The trajectory never crosses zero, implying that the cross section
grows in the physical region with increasing $\sqrt{-t}$.

We note that the trajectories decrease rapidly with increasing
$\sqrt{-t}$, showing that most of the variation is located around 0.
The induced trajectories are sensitive to the instanton size
through a rescaling of the induced form-factor by the parameter
${\bf a}$. Indeed, the slope $\Delta' (0)$ 
relates to the pomeron slope through,
\be
\Delta'(0)=\alpha' = (0.5-0.2)/{\rm GeV}^{2}\,\,,
\ee
with 0.5 corresponding to the unmodified instanton induced form
factor (${\bf a}=0$) and 0.2 corresponding to a scaled down instanton
induced form factor (${\bf a}=0.25$).

Our main point is that the smallness of the soft pomeron intercept 
$\alpha'$ reflects directly on the smallness of the squared instanton 
radius.  Our trajectory
curves are similar to those reported by KKL~\cite{KKL} (who also
continued their trajectories to the unphysical region $t>0$).
However, this comparison is only qualitative,
 since the induced form factor used in KKL differs
fundamentally from the one we have derived (see discussion below).
The issue of the size dependence was not addressed in~\cite{KKL}.
\begin{figure}[h]
\vskip 0.2in
\includegraphics[width=2.9in]{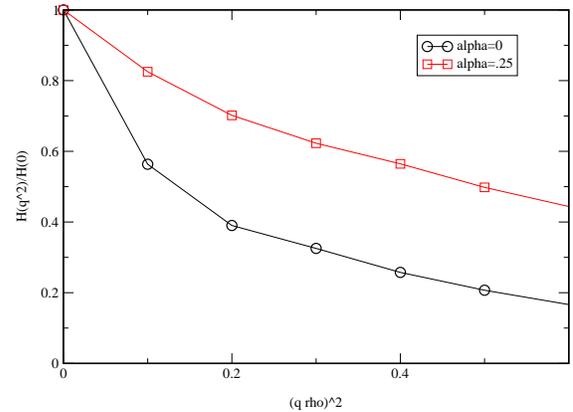}
\vskip 0.2in
\caption[]{
 \label{fig_H}
Instanton-induced form factor ${\bf H}(q_\perp^2)$ at the origin of the soft
Pomeron trajectory, normalized to its value at $t=0$, versus 
$-t\rho_0^2=q_\perp^2\rho_0^2$. The squares refer to the unmodified 
instanton shape and the circles to the modified one.}
\end{figure}

\subsection{Instanton Shape Dependence}
\label{shape}
An ensemble of instantons in the QCD vacuum  is always described by
retaining the instanton configurations in the singular gauge, where
the topological singularity is located at the instanton center\footnote{This
is in
contrast to the regular gauge where the singularity is removed to
the sphere at infinity. Some subtleties regarding this transformation
are discussed in \cite{YZ}.}. In the singular gauge, the instanton
gauge field falls off as $A\approx \rho_0^2/x^3$ at large $x$,
providing a ground for a dilute analysis.
 By keeping the topological properties at the center, a 
modification of the ``tail'' of the 
gauge field at large distances may be allowed. Moreover, 
since the semi-classical
analysis holds only for strong fields, these tail modifications 
are in general expected.

There are many effects which can modify the tail of the instantons,
an example is through interactions. Indeed an early variational 
estimate of the instanton vacuum energy using exponentially modified 
instanton form factors such as $e^{-{\bf a}\, x/\rho_0}$~\cite{DP} 
show a minimum at ${\bf a}\approx 0.5$. Other possible reasons for 
instanton-shape modifications can be due to confinement as discussed
in the context of the QCD dual superconductor~\cite{Shur_sizes}. Lattice
studies of various  gluonic correlators also show rapid exponential 
fall-off. Thinking about the lowest glueballs, with their 2 GeV mass 
scales as 2-gluon bound state, may imply an even larger sizing down
with  ${\bf a}\approx 1$.

Interestingly enough, the phenomenological issue regarding the instanton shape
has never been seriously addressed. It is because (to our
knowledge) all previous applications of instanton physics in QCD  were
found to be generally insensitive to it. Indeed, most applications
(see \cite{SS_98}) are related to light quarks and fermionic
zero modes, which exist and are normalized independently of the instanton
field. Also, the  correlators of the scalar field strength
combination $G_{\mu\nu}^2$~\cite{SS_glue}
are instanton-shape insensitive since they fall-off rapidly at
large $x$ (as $1/x^8$ in perturbation theory). This is
in sharp contrast to the present case, where the induced form factors
fall-off as $1/x_\perp^2$ and their integrated effect 
in the transverse plane (through $dx_\perp$) diverges 
logarithmically.

Instanton-shape modifications cause our results 
(intercept and slope) to change quantitativaly, 
showing the limitations in the present analysis.
To illustrate this, consider changing the description from
regular to singular gauge and inserting the exponential modifications
in the instanton tails. Hence, the argument in the definition 
of the induced instanton form factor ${\bf J}$ changes accordingly
\be
\frac {\pi\,
|x|}{\sqrt{x^2+\rho_0^2}} \rightarrow \pi\left( \frac {
|x|}{\sqrt{x^2+\rho_0^2}}-1\right)\,\,e^{-{{\bf a}\,|x|/\rho_0}}\,\,.
\label{alpha}
\ee
The effect on the form-factor {\bf J} is illustrated in
Fig.~\ref{fig_J}. In the unmodified case with ${\bf a}=0$ (top points) 
the form factor rises towards small momentum transfer, which is not
the case for ${\bf a}>0$. As expected the modifications due to ${\bf a}$
are small at large $q_\perp^2$.

\subsection{Reggeization}

Hadrons are usually composed of several partons, each 
capable of a pair-collision in the transverse plane.
Even with only 3 quarks in a nucleon (ignoring
``primordial'' glue etc), we have $3\times3=9$
possible sub-collisions in the NN case. Their probability 
depends on the quark distribution in the transverse plane which
is given by some pertinent light-cone wave functions. The latters 
will not be important for the qualitative arguments to follow.
The chief  argument is that since the hadron diffractive size $R_H$ 
($R_H\approx 3r_0\approx 1\,{\rm fm}$) is large in comparison  
to the interaction range,   the maximum number of  
independent collisions can be as large as
$N_*\approx R_H^2/\rho_0^2\approx 10\gg 1$. If we  treat those
pair-collisions as statistically independent (Poisson) and proceed
to resum their respective probabilities $\sigma/\pi\rho_0^2$, the
total hadron-hadron cross section $\sigma_{HH}$ is
\be
{\sigma_{HH} (s,t)}\approx \,{\pi\, R_H^2} \, \sum_{n=1}^{N_*} \,
\frac 1{n!}\,
\left(\frac {\sigma (s, t)}{\pi\rho_0^2}\right)^n\,\,, 
\label{POM4}
\ee
which is a polynomial in ${\rm ln}\,s$ of degree $N_*\approx 10$.
Since (\ref{POM4}) is a  polynomial in $\kappa_0\approx 0.01$ 
as well, the expansion is well approximated by its two
first terms within the Froissart bound. Note that
by considering $N_*$ to be infinite, we obtain 
a reggeized cross section ($\rho_*\approx \rho_0$)
\be
\sigma_{HH} (s,t) \approx \,\pi\,R_H^2\,
\left( e^{\alpha_s/\pi}\,\,s^{\Delta (t)} -1\right) \,\,,
\label{POM44}
\ee
with an asymptote $s^{\Delta (t)}$ fixed by the instanton density.
We now proceed to compare our analysis and results to recent suggestions,
as well as discuss the limitations of some of our assumptions.

\section{Weak-Field Approximation}
\label{sec_ff}

In this section we will provide some qualitative arguments regarding
the relationship between the semi-classical analysis carried above
and the weak-field approximation used in previous analyses, e.g.
in KKL~\cite{KKL}. 

\subsection{Weizsacker-Williams Approximation}

Consider that each hard parton is surrounded by a
cloud of wee partons making a virtual Coulomb field in its rest frame. 
A hard parton with large rapidity $y$ (not to be confused with
a y-coordinate in this section) going through a classical field, 
radiates quasi-real gluons
\be
Q(p) \rightarrow Q(k) + g^* (q)
\label{WF1}
\ee
with a Weizsacker-Williams distribution
\be
dN_{WW} \approx \frac {\alpha_s}{\pi}\,\frac{d\omega}{\omega}\,
\frac {dq_{\perp}^2}{q_\perp^2}\,\,{\bf F}_E^2 (q_\perp^2)\,\,,
\label{WF2}
\ee
for fixed energy $\omega=q_0$ and transverse momentum. Terms of order
$\omega/\sqrt{s}$ and $q_\perp^2/s$ have been ignored. ${\bf F}_E$ is the
color-electric Sachs form-factor of the hard parton induced by the
classical field, which is 1 for a point charge.

Hard parton-parton scattering in the Weizsacker-Williams approximation
follows in two stages
\be
Q(p_1) +Q(p_2) \rightarrow &&
Q(k_1) + Q(k_2) \nonumber\\
&& + \left(g^* (q_1) + g^* (q_2)\rightarrow {\bf X} (q_1+q_2)\right)\,\,,
\label{WF3}
\ee
from which the inelastic cross section follows by convoluting the
entrance fluxes (\ref{WF1}) with the gluon-gluon fusion cross section.
Setting $dy=d\omega/\omega$ we have
\be
d\sigma \approx  &&
\left(\frac{\alpha_s}{\pi}\right)^2\, 
{dq_{1\perp}^2 \over q_{1\perp}^2}{d{q}_{2\perp}^2 \over
{q}_{2\perp}^2}\,dy_1\,dy_2\nonumber\\
&&\times\, {\bf F}_E^2 ( q_{1\perp}^2 ) \,{\bf F}_E^2 ({q}_{2\perp}^2)
\,\,\sigma_{gg}(q_{1\perp} -q_{2\perp},y_1-y_2)\,\,.
\label{WF4}
\ee
The total fusion cross section sums over all the exclusive cross
sections in (\ref{WF3}). In particular, it depends only on the rapidity 
difference (the energy of the $gg$ sub-process) and the transferred
transverse  momentum. 
Thus, one can integrate over the CM rapidity $Y=(y_1+y_2)/2$, which is
bracketed by the rapidity of the original hard partons, leading to
the standard ${\rm ln}\,s$ enhancement. The logarithmic rise in the
inelastic cross section is just a measure of the available longitudinal
phase space of the produced subsystem. The magnitude of the rise depends
on the exclusive cross section $\sigma_{gg\rightarrow {\bf X}}$ summed over
all final states ${\bf X}$, and the induced form factors that we now discuss.

\subsection{Induced Form-Factors}

The instanton induced form factors are important elements of the
high-energy scattering calculations we have described. They make 
all transfer integrations finite, thereby determining the magnitude
of the cross section. They also keep the instanton effects from being 
part of the hard processes. In a recent investigation by KKL~\cite{KKL},
the instanton induced form factor (now in absolute units) 
\be
{\bf F}_E(q_\perp^2)\approx 
\frac 1 {(q_\perp \rho_0)^4} 
\left(1- \frac 12\,{\,(q_\perp\,\rho_0)^2}\ K_2 (q_\perp \rho_0)\right)\,\,,
\label{WF5}
\ee
was derived using the weak-field limit. (\ref{WF5}) 
is simply   the Fourier transform of the instanton field
(\ref{WE5x}).  The weak-field approximation is justified
if only a single-gluon exchange between the through-going
parton and the instanton is registered. The single-gluon
approximation is justified when the parton impact parameter exceeds the instanton
size $\rho_0$, which is equivalent to $q_\perp\,\rho_0 \ll 1$. 
But in this case, the form factor (\ref{WF5}) reduces to 1.

In general, the parton-instanton interaction  takes place when 
the incoming parton punches through the instanton at an impact 
parameter comparable to the instanton size $\rho_0$, for which
$q_{\perp}\, \rho_0\leq 1$. Hence, we cannot use the weak
field approximation and the induced form factors are given by
Wilson-lines
\be
{\bf F}_E (q_\perp^2) \approx 
\int \frac {d^3 x}{\rho_0^3} \,e^{iq_\perp\cdot x} 
\,({\bf W}(\infty, x_\perp, x_3) -{\bf 1})\,\,,
\label{WF6}
\ee
with open color indices (see above). Our induced form
factors resum multiple gluon exchanges, in contrast to those
discussed by KKL~\cite{KKL}.

\section{Additional Diffractive Contributions}

Throughout two semiclassical resummations were used: (i)
the eikonalized phases which resum multiple interaction with
the through-going partons, and (ii) the produced
gluons which resum into an instanton-antiinstanton interaction. 
In this section, we discuss additional effects that we have
not retained.

\begin{figure}[t]
\vskip 0.2in
\begin{center}
\includegraphics[width=1.5in]{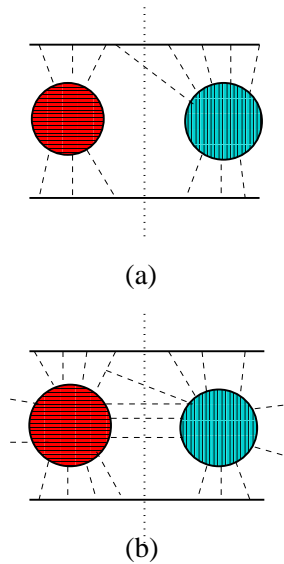}
\end{center}
\vskip 0.2in
\caption[]{
 \label{fig_notincluded}
The interference (a) and interaction (b) diagrams not included in
our analysis.}
\end{figure}

\subsection{Interference and Interaction }

The diagrams shown in 
Fig.~\ref{fig_notincluded} describe additional interference (a)
and interaction (b) effects between the gluonic radiation 
from the hard partons and the instanton, which we have not
considered in our analysis.
The interference of the radiation (a) with the instantons can be
argued to be small, for the following reason. Kinematically 
the radiation from the external line has a flat rapidity 
distribution extending all the way to the rapidity of the
hard parton (see the Weizsacker-Williams approximation),
while instantons produce about $S_0\approx 10-15$ gluons 
within a cluster occupying one unit in rapidity space.
The overlap between the radiation and the 10-15 gluons is
about $1/{\rm ln}\,s$ thereby compensating the logarithmic
growth in the cross section.

A possible way to include the radiation effects is to
calculate the eikonal factors ${\bf W}$ in the {\it combined}
field of an instanton-antiinstanton in the simplest sum ansatz.
The interaction diagrams (b) can then be viewed as additional
corrections to this simple ansatz, diagrammatically describing 
a more appropriate solution. Since no analytical
formulae for path-order exponents in a field more complicated than that of a
single instanton is available, inclusion of those corrections would
complicate the present analysis considerably. We hope to report on these
effects elsewhere.

\subsection{Additional Partons}

As we did emphasize at the beginning of this paper, our analysis of
the collision processes is based on the parton model, whereby the 
{parton-parton} cross sections (evaluated) are separated from the
{hadronic wave functions} (not evaluated).  The separation 
depends on what is exactly meant by our distinction of a { parton} 
from {through-going quark}, which is of course scale dependent.
In our case, the separation scale is set by the mean instanton size
$\rho_0\approx 1/(600 \, {\rm MeV}) \approx 1/3 \, {\rm fm}$. This choice
may appear to be in contradiction with the usual statement
that perturbative
QCD cannot be used
below the scale  of 1 GeV. However, there is no contradiction
if those deviations from perturbative QCD are precisely due to
the instanton effects we account for. For phenomenological and theoretical
arguments in favor of this view-point in the vacuum we refer to
\cite{SS_98,RRS}. Our present analysis, extends these arguments to
diffractive scattering. Although we have ignored perturbative gluons 
and field-theoretical renormalizations of all quantities discussed, 
we believe they have to be included around our semi-classical treatment
for our results to be complete.

At the low normalization scale of order $0.6$ GeV the partons are 
{dressed}  by their surrounding fields, perturbative and
non-perturbative, and for all purposes are effective objects.
This  is why we refered above to the nucleon as being made 
of 3 quarks, the photon of a quark-anti-quark, etc.
Most of the glue and the sea quarks and 
antiquarks seen in DIS structure functions  reside inside those 
effective objects. The simplest part of the dressings are ``wee''
partons included above in the {\bf W} factors. These are mostly
the perturbative fields surrounding the parton\footnote{ In the
eikonalized form factors only the lowest coupling through $g$ was
retained. This is supported by experimental data from HERA, showing that
at the low normalization point $Q^2\approx 1\,{\rm GeV}^2$, the gluonic density
is about flat (even decreasing) towards small $x$. 
If there is a need to describe hard processes, then additional
diagrams as for instance those retained in the DGLAP evolution, should
be included to reach higher normalization points and account for the
growing gluon density towards small $x$.}. Still,
 not all parton dressing is perturbative.
Indeed, in the simplest constituent quark model for hadron
spectroscopy the effective wave-function for $\bar q q$ and $qqq$
still requires the help of a confining potential, or a QCD string.
Therefore our results should be improved by including the corresponding
effects of light-cone wave-functions.

Finally, phenomenological structure functions,
even taken at low normalization points, do exhibit a non-zero gluon 
density. This suggests that there are also {through-going gluons}
inside the high energy nucleon or photon\footnote{In  the 
stochastic vacuum model \cite{DOSCH}, those gluon-exchanges 
originating from the canvassing strings are assumed to
be even dominant in 
the hadronic cross section.}, for which the instanton-induced cross 
section has not been assessed yet. There is no technical difficulty 
to do so along the lines we have presented, and we expect their cross 
section to be generally  larger, improving the agreement between 
our theoretical prediction and empirical rate of the  cross section growth.

\section{Conclusions and Outlook}

In this work we have extended our evaluation of the  instanton 
contribution to the scattering of partons at large $\sqrt{s}$ and 
small $-t/s$, from quasi-elastic to inelastic collisions. 
The present findings and results support our
original analysis~\cite{sz00} stressing the importance of 
instantons in high-energy and near-forward scattering
amplitudes in QCD. Although our analysis differs significantly 
from the one recently reported by KKL in~\cite{KKL}, the underlying 
physics is about the same. All in all, instantons are shown to play
a significant role in diffractive processes.

Throughout, we have tried to make a consistent use of 
purely semiclassical treatments.
The interactions with the through-going partons are included
in the eikonalized factors, with  any number of 
gluons, and the inelastic production of any number of gluons 
is summed into an instanton-anti-instanton interaction.
Both approaches have been developed previously, but their
combined application to the soft pomeron problem is new.
At higher invariant masses of the produced system, we also carried out 
unitarity considerations (through a chain of alternating instanton 
and antiinstanton) in somewhat more details than it has been done previously.

Major differences between quasi-elastic and inelastic processes have
been found. (i) The production of an intermediate 
multi-gluon system leads to a ${\rm ln}\,s$ growth in the total 
cross section, opening up the possibility to explain soft pomeron 
physics after pertinent resummation. (ii) The inelastic cross section is 
much larger, in fact it is parametrically larger by an inverse
power of the instanton diluteness parameter (about a factor of 100). 
The pomeron  intercept is small because it is simply proportional 
to the {\em small instanton diluteness} parameter in the QCD vacuum. 
(iii) The pomeron
slope is small,  because it is  directly related to the instanton induced 
form-factors on the eikonalized hard partons, hence to {\em small instanton
sizes} in the QCD vacuum. (iv) For the first time in QCD applications of
instantons,  we have found that  the instanton shape at large
distances from the center can actually impact on a physical
observable such as the pomeron slope.

Our work can be extended in a number of ways.
The most straightforward extension is to small size
dipoles~\cite{sz00} and gluons as partons participating in scattering.
Small size dipole scattering is related to processes 
with virtual photons, such as $\gamma^*\,h$ and $ \gamma^*\,\gamma$
and even  $\gamma^*\,\gamma^*$ with two virtual 
photons. The next set of questions which can be also addressed
in the present framework relates to the nature of the produced 
multi-gluon systems in the exclusive reactions, or the ``sphaleron decay''
problem. The total cross section we evaluated can be decomposed into 
pertinent channels with given quantum numbers involving specific hadrons and
glueballs.  Since in the inelastic processes, particle production
is in general {masked} by multiple production from string decays 
from the final stage, we suggest to focus on {double diffractive}
cross sections where the produced hadrons are selected alone and 
separated by large rapidity gaps from the target and the projectile.

Finally, one may ask what happens at very large energies.
We have briefly discussed a semiclassical resummation of our basic
process, resulting in a cross section growing as ${\rm ln}\,{s}$. 
The total hadron-hadron cross section resulted from a Poissonian 
resummation of pair-independent parton-parton scatterings or {
bushes}.  We have not addressed the issue of rescattering between
the produced gluons as in the {instanton ladders} considered
in KKL~\cite{KKL}, although this can be explored beyond the
semi-classical framework we have detailed. Clearly, a thorough 
comparison of the present ideas and results with the data will
eventually tells us if there is room for hypothetical { primordial}
small $x$ partons, distinct from the usual dressing of valence quarks 
and residing in different positions in the transverse plane.\\\\

{\bf Acknowledgments}
\\
This work was supported in parts by the US-DOE grant
DE-FG-88ER40388 and by the Polish Government Project (KBN) 2P03B 00814.

\end{document}